\documentclass[prd,superscriptaddress,amsfonts,amssymb,amsmath,showpacs,twocolumn]{revtex4}
  
    \usepackage{epsfig}
    \usepackage{graphics}  
   \usepackage{color}
   \usepackage{amsmath}
    \date{\today}

\begin{document}

 \title{Traces of the Unruh effect in surface waves} 

    \author{Guilherme B. Barros}\email{gb.barros@unesp.br}    
    \affiliation{Instituto de F\'\i sica Te\'orica, Universidade
    Estadual Paulista, Rua Dr.\ Bento Teobaldo Ferraz, 271, 01140-070,
    S\~ao Paulo, S\~ao Paulo, Brazil}
    
      \author{Jo\~ao P. C. R. Rodrigues }\email{jp.rodrigues@unesp.br}    
    \affiliation{Instituto de F\'\i sica Te\'orica, Universidade
    Estadual Paulista, Rua Dr.\ Bento Teobaldo Ferraz, 271, 01140-070,
    S\~ao Paulo, S\~ao Paulo, Brazil}
    
    \author{Andr\'e G.\ S.\ Landulfo}\email{andre.landulfo@ufabc.edu.br}
    \affiliation{Centro de Ci\^encias Naturais e Humanas,
    Universidade Federal do ABC,
    Avenida dos Estados, 5001, 09210-580,
    Santo Andr\'e, S\~ao Paulo, Brazil}
 
    \author{George E.\ A.\ Matsas}\email{george.matsas@unesp.br}
    \affiliation{Instituto de F\'\i sica Te\'orica, Universidade
    Estadual Paulista, Rua Dr.\ Bento Teobaldo Ferraz, 271, 01140-070,
    S\~ao Paulo, S\~ao Paulo, Brazil}


\begin{abstract}

We look for classical traces of the Unruh effect in gravity waves. For this purpose, we start considering a 
white noise state of gravity waves on the surface of a water basin and calculate the two- and four-point 
functions of the Fourier transform of the surface-height field with respect to accelerated observers. The 
influence of the basin boundaries and possible deviations from Gaussianity in the white noise state are 
considered in order to approximate conditions attainable in the laboratory. Eventually,  we make the basin 
infinitely large in order to make contact between our classical results and quantum ones derived in free space. 
We hope that our results help to strengthen the bridge between the Unruh effect and this classical analog.

\end{abstract}

\maketitle

\section{Introduction}
\label{sec:I}

According to the Unruh effect,  Rindler observers, i.e. uniformly accelerated observers in the Minkowski 
spacetime, associate a thermal bath to the usual no-particle state as defined by inertial observers 
(Minkowski vacuum). The temperature of the Unruh thermal bath as measured by Rindler observers
 with proper acceleration $a$ is given by~\cite{unruh1976notes}
\begin{equation} \label{unruhtemperature}
    T_U = \frac{\hbar a}{2\pi k_B c}.
\end{equation}
It is not easy to directly observe the Unruh temperature with {\it present} technology, 
although feasible proposals can be found in recent literature \cite{PhysRevLett.118.161102}. This
can be easily seen from Eq.~(\ref{unruhtemperature}), since an acceleration of about $ 10^{20} \, {\rm m/s^2}$ 
would be needed in order to reach an Unruh temperature of $1 \, {\rm K}$. Faced with this situation,
one may wonder whether some analog of the Unruh effect could be seen in some condensed-matter 
system. This seems promising because of the following two main features: 
\begin{enumerate}
    \item The speed of light, $c$, would be replaced in Eq.~(\ref{unruhtemperature}) 
    by the speed of the phonon, quasi-particle or other medium perturbation, $v \ll c$, 
    enhancing the Unruh temperature by a huge factor. In Bose-Einstein condensates~\cite{HZC19}, 
    e.g., $v \sim 1 \, { \rm mm/s}$ \cite{SoundBE}, increasing the Unruh temperature 
    by a factor of $10^{11}$;
    \item The proper acceleration $a$ would be replaced by an analog proper acceleration 
    $A$ \cite{VisserReview}. It turns out to be much easier to imprint a large analog acceleration 
    $A$ rather than a large physical acceleration $a$ to an observer, leading to an extra
     enhancement to the analog Unruh temperature. 
\end{enumerate}
Analog models can be both classical or quantum. Here, we will be interested in classical analogs
of the Unruh effect because classical phenomena occur at usual scales of length and time, making 
experiments more feasible. For a general discussion on the Unruh effect in classical field theory, see, 
e.g., Ref.~\cite{HM93} and for a specif application to classical electrodynamics, see Ref.~\cite{LFM19}.

Interestingly enough,  Leonhardt et al. have recently looked for traces of the Unruh effect in gravity waves present 
on the surface of a one-dimensional water basin~\cite{Leonhardt2018}. They have shown how an observer evolving in 
a Gaussian white noise with analog proper acceleration $ A  = {\rm const} $ can read an analog Unruh temperature, 
$$
T_A = \frac{\hbar A}{2 \pi  k_B v},
$$ 
from the two-point function in momentum space calculated in its proper frame, where $v$ is the 
gravity-wave propagation speed. In their analysis, they consider the basin long enough in order to 
ignore boundary effects. It seems, thus, necessary to complement this investigation wondering how 
the presence of boundary conditions can impact the laboratory outputs.  In addition, we analyze how 
deviations from Gaussianity may impact higher-order point functions by looking at the 
four-point  function.  For the sake of consistency, we check that our results lead to the usual Unruh effect when 
no boundaries are present.

This paper is organized as follows: in Sec.~\ref{qvacuum}, we discuss what are the main properties of the quantum 
vacuum, which should be considered in our classical analog system. In Sec.~\ref{Sec:AnalogVac}, we introduce the analog 
spacetime. In Sec.~\ref{sec:2point}, we show how boundary conditions affect the two-point function extracted by 
accelerated observers. In Sec.~\ref{sec:Unruheffect}, we establish a direct connection between the Unruh effect and 
Sec.~\ref{sec:2point} results. In Sec.~\ref{sec:othernoises}, we discuss the impact of different choices of white noise on the 
four-point function. Our final comments appear in section~\ref{sec:conclusions}. We adopt metric signature $(-,+,+,+)$.
We keep  $c$ and $\hbar$ in our formulas in order to make easier the comparison of the results coming 
from the full Unruh effect  with the corresponding ones coming from this nonrelativistic classical analog.  

\section{Quantum Vacuum: essential features} \label{qvacuum}

In this section, we briefly review some properties of the quantum vacuum that will be essential to our problem. 
We start by considering a free real massless scalar field $\hat{\Phi} (x^\mu)$ in the spacetime 
$$
(\mathbb{R}\times [-L/2,L/2],\eta), \quad L={\rm const},
$$ 
endowed with a Minkowski metric $ \eta $. We have chosen such a spacetime because, after all, any real experiment
takes place in a compact domain. Let us cover it with Cartesian coordinates $\{x^{\mu}\}=\{t,z\}$, $|z| \leq L/2$. 

Now, let us expand $ \hat{\Phi} (x^\mu)$ in terms of a complete set of normal modes satisfying periodic boundary conditions
and orthonormalized by the Klein-Gordon inner product, as usually: 
\begin{equation}
    \hat{\Phi}(x^{\mu}) = 
    \sum_{\substack{m=-\infty \\ m \neq 0}}^{\infty} 
    \left( {\frac{\hbar c^2}{2\varpi_{{m}}L}} \right)^{1/2}  
    [\hat{a}_{{m}} e^{i k_{\mu} x^{\mu}} + \hat{a}^{\dagger}_{{m}} e^{-i k_{\mu} x^{\mu}}],
    \label{equ:quantfield}
\end{equation} 
where 
$k^{\mu}=(\varpi_m/c, k_m)$ 
with 
$$
\varpi_m = |k_m| c \quad {\rm  and}  \quad k_m = 2m\pi/L.
$$ 
The Minkowski vacuum $|0 \rangle$ is defined by imposing 
$ \hat{a}_{{m}} |0 \rangle \equiv 0 $ 
for all 
$ {m}$. The canonical commutation relations between $\Phi(x^\mu)$ and its conjugate momentum $\Pi (x^\mu)$ 
leads to $ [\hat{a}_{{m}}, \hat{a}^{\dagger}_{{m'}}] = \delta_{{m}{m'}}$. 

Now, let us write $ \hat{a}_{{m}}$ in terms of the Hermitian operators 
$ \hat{q}_{{m}}$ and $\hat{p}_{{m}}$ as
\begin{equation}
    \hat{a}_{{m}} \equiv (\hat{q}_{{m}} + i \hat{p}_{{m}})/\sqrt{2}.
    \label{equ:aoperator}
\end{equation}
 A thorough check shows that any-order correlation functions  for $\hat{q}_m$ and $\hat{p}_{m'}$  are those associated with 
 a Gaussian distribution:
\begin{eqnarray}
 \Psi (q_m)  & \equiv &    \langle q _m | 0 \rangle = \frac{1}{\pi^{1/4}} e^{-q_m^2 /2} ,
 \label{equ:psi} \\ 
  \Phi (p_m)  & \equiv &    \langle p _m | 0 \rangle = \frac{1}{\pi^{1/4}} e^{-p_m^2 /2} .
 \label{equ:phi} 
 \end{eqnarray}
 In particular, the ``first-'' and second-order correlation functions are
\begin{eqnarray}
 \langle 0 | \hat{q}_{{m}} | 0 \rangle 
= 
\langle 0 | \hat{p}_{{m}} | 0\rangle 
&=&
0 
\label{equ:corr1} 
\\
\langle 0 | \hat{q}_{{m}} \hat{q}_{{m'}} |0 \rangle 
= 
\langle 0 | \hat{p}_{{m}} \hat{p}_{{m'}} |0 \rangle 
&=& 
\delta_{{m}{m'}}/2 
\label{equ:corr2}
\\
(1/2)\langle 0 |\hat{q}_{{m}} \hat{p}_{{m'}} + \hat{p}_{{m'}} \hat{q}_{{m}} |0 \rangle 
&=& 0 
\label{equ:corr3}
\end{eqnarray}
 (The left-hand side of Eq.~(\ref{equ:corr3}) was defined from averaging between 
expressions which lead to the same classical quantity.) 

This is the Gaussian nature of the quantum vacuum, which we must bring into the classical state.

\section{A classical analog of the Minkowski vacuum} \label{Sec:AnalogVac}

In order to establish a bona fide classical analog of the Minkowski vacuum, we begin by considering a 
perturbation $\mathcal{A}$ on the surface of a water basin of length $L$ and depth $h$. 
The system is assumed to be in the Galileo spacetime, \textit{i.e.}, the spacetime of classical mechanics, 
which will be also covered with Cartesian coordinates $\{x^{\mu}\}=\{t,z\}$, $|z| \leq L/2$. The waves  
are restricted to propagate only in one spatial  dimension. Besides,  it is assumed that (i)~$\mathcal{A}\ll h$
and (ii)~$|\partial \mathcal{A}/ \partial t| $ is much smaller than any other velocity scale 
in the problem. 

Then, it is possible to show that an arbitrary perturbation on the water surface can be written as (for more 
details see, e.g., Eqs.~(3)-(5) in Chap.~IX of Ref.~\cite{lamb2015hydrodynamics}) 
\begin{equation} \label{Afield0}
\mathcal{A}(t,z) 
=  \sum_{\substack{m=- \infty \\ m \neq 0}}^{+\infty} [ {\cal A}_m (t,z) + {\cal A}^*_m (t,z) ] , 
\end{equation} 
where
\begin{equation}
{\cal A}_m (t,z)=
\alpha_m \sqrt{\frac{v_m}{2 \varpi_m L}}  \cos \left[ k_m \left( z+ \frac{L}{2} \right) \right]   e^{-i \varpi_m t} 
\end{equation}
satisfies
\begin{equation}\label{modesequation}
 \frac{\partial^2 \mathcal{A}_m}{\partial z ^2} - 
\frac{1}{v_m^2}\frac{\partial^2 \mathcal{A}_m}{\partial t^2}=0   
\end{equation}
with boundary conditions
\begin{equation}\label{modesboundaryconditions}
 \left. \frac{\partial \mathcal{A}_m}{\partial z}\right|_{\substack{z=\pm L/2}}=0.
\end{equation} 
Here, $\alpha_m = {\rm const} \in \mathbb{C}$, 
$$
v_m = \sqrt{(g/k_m)\tanh(k_mh)}, \quad g\approx 9.8 \, {\rm m/s^2},
$$  
$$
\varpi_m \equiv |k_m| v_m, \quad k_m\equiv m\pi/L.
$$   
We found it convenient to keep ${\cal A}_m$ and $\alpha_m$ with the same unit (of length)
in contrast to Leonhardt et al~(see Eqs.~(12)-(13) of Ref.~\cite{Leonhardt2018}).  
 
In order to avoid dispersion, our perturbation ${\cal A}(t,z)$ will be assumed to be a superposition of 
modes satisfying 
\begin{equation}\label{equ:condition}
|k_m h| = |m|\pi h/L \ll 1, 
\end{equation}
in which case
\begin{equation}
v_m \approx v\equiv \sqrt{gh}.
\label{velocidade}
\end{equation}
Physically, this means that we will be coarse-graining time intervals of order
$ (h/10~{\rm m})^{1/2} \;{\rm s}$. 

It is worthwhile to emphasize that our system will be classical under any realistic conditions.
This can be seen from  
$$
\frac{{\cal A}_m {\cal P}_m}{\hbar}  
\gg  10^{30}  
\left( \frac{{\cal A}_m}{1\; {\rm cm}} \right)^{5/2}
\left( \frac{x_\perp}{10\; {\rm cm}} \right)^{2} 
\left( \frac{\rho}{\rho_{\rm H_2 O}} \right),
$$
where 
$$
{\cal P}_m = (v^2/\varpi_m) \,h \, x_\perp \, \rho
$$ 
is the momentum corresponding to 
mode ${\cal A}_m$~(see p. 419 of Ref.~\cite{lamb2015hydrodynamics}), 
$x_\perp \ll v/\varpi_m$ is the length scale of the spatial direction perpendicular to 
the wave propagation,  and $\rho$ is the fluid density. 

Under assumptions~(\ref{equ:condition})-(\ref{velocidade}), Eq.~(\ref{Afield0}) 
becomes
\begin{eqnarray} \label{Afield}
\mathcal{A}(t,z) 
&=& \sum_{\substack{m=-N \\ m \neq 0}}^{N} \sqrt{\frac{v}{2 \varpi_m L}}\cos \left[ k_m \left( z+ \frac{L}{2} \right) \right]  
\nonumber \\
&\times &[\alpha_m e^{-i \varpi_m t} + \alpha_m^{*} e^{i \varpi_m t}],
\end{eqnarray} 
where the summation must be restricted to some $N\ll L/(\pi h)$.
As a consequence, $\mathcal{A}(t,z)$ will satisfy 
\begin{equation*}
  \frac{\partial^2 \mathcal{A}}{\partial z^2} - \frac{1}{v^2}\frac{\partial^2 \mathcal{A}}{\partial t^2} =0,
  \label{equ:wave}
\end{equation*}
which can be cast in the covariant form
\begin{equation*}  \label{waveequation}
\Box \mathcal{A}(x^{\mu}) = 0, \quad \Box = g^{\mu \nu}\partial_{\mu}\partial_{\nu}.
\end{equation*}
Here, $g_{\mu\nu}$ is the associated gravity-wave metric, which endows the analog Minkowski  spacetime 
$$
(\mathbb{R}\times [-L/2,L/2], g).
$$
Its components in Cartesian coordinates $\{t,z\}$ can be read  from
 \begin{equation} \label{macustica}
    ds^2 = g_{\mu\nu}dx^{\mu}dx^{\nu} =  -v^2dt^2 + dz^2.
\end{equation}  

Now, with the purpose of bringing the desired aspects of the quantum vacuum to the classical world, 
we make use of the only parameters in the field $\mathcal{A}(t,z)$ that are not determined by the 
laws of hydrodynamics, but rather by the initial conditions of the system: the complex coefficients 
$ \alpha_m $. Inspired by the Sec.~\ref{qvacuum} discussion, we define
\begin{equation*}
    \alpha_m \equiv \frac{1}{\sqrt{2}}(q_m + i p_m), \quad q_m,p_m \in \mathbb{R},
\end{equation*} 
and choose $ q_m$ and $ p_m$ to be Gaussian random variables according to the rules laid out by 
Ref.~\cite{Leonhardt2018}: for each mode $ m $, they will be randomly chosen from the 
uncorrelated Gaussian probability distribution function 
\begin{equation} \label{gauss}
    P(q_m) = \frac{1}{\sqrt{\pi I}} e^{-q_m^2/I},\; P(p_m) =\frac{1}{\sqrt{\pi I}} e^{-p_m^2/I}.
\end{equation} 
It can be shown that
\begin{eqnarray}
    \left\langle q_m \right\rangle = \left\langle p_m \right\rangle &=&0, \label{equ:clascorr1} \\
    \left\langle q_m q_{m'} \right\rangle = \left\langle p_m p_{m'} \right\rangle &=& I \delta_{mm'} /2, \label{equ:clascorr2}\\
    \left\langle q_m p_{m'} \right\rangle &=& 0, \label{equ:clascorr3}
\end{eqnarray} 
for all $m$, $m'$.  Equation~(\ref{equ:clascorr2}) 
has an extra real constant $I$ with unit of squared length in comparison to Eq.~(\ref{equ:corr2}), 
giving the strength of the classical correlation. Equation~(\ref{equ:clascorr3}), for its turn, 
should be seen as the classical version of Eq.~(\ref{equ:corr3}), where $ \hat{q}_m \hat{p}_{m'} $ and 
$\hat{p}_{m'} \hat{q}_{m} $, corresponding to the same classical quantity, were averaged out. From now on, 
every time distinct quantum operators lead to the same classical function, we will repeat the same
procedure as in Eq.~(\ref{equ:corr3}). 

In order to see how the choice of $P(r_m)$, $r_m=q_m, p_m$, impacts on the correlation functions, we will compare the results obtained 
when one chooses $q_m$ and $p_m$  from the uncorrelated  Gaussian probability distribution~(\ref{gauss}) against 
the ones obtained when we choose $q_m$ and $p_m$  from the uncorrelated  uniform probability distribution:
\begin{equation} \label{uniform}
 P(q_m) \! = \! \frac{H (\sqrt{3I/2}-|q_m|)}{\sqrt{6I}} ,  P(p_m) \!=\! \frac{H (\sqrt{3I/2}-|p_m|)}{\sqrt{6I}},
\end{equation} 
where $H(x)$ is the Heaviside function. We emphasize that although the Gaussian and uniform distributions 
above lead to the same first and second momenta~(\ref{equ:clascorr1})-(\ref{equ:clascorr3}), 
they will not lead to the same higher-order ones. We will have more to say about it in Sec.~V. For now, it is 
enough to say that the closer to the Gaussian distribution~(\ref{gauss}), the better 
classical state  one has to mimic the quantum vacuum.

\section{Boundary effects on the two-point functions} \label{sec:2point}

\begin{figure}
   \centering
   \includegraphics[width=87mm]{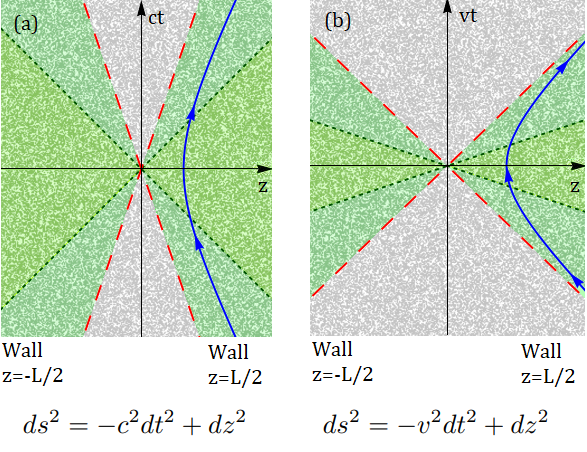}
    \caption{The solid line depicts the worldline of a uniformly accelerated observer in both the 
    (a)~Minkowski  and (b)~analog spacetimes. The dotted line represents the light 
    cone, while the dashed line represents the analog cone (associated with gravity waves 
    moving with speed $v$ in the geometrical optics limit). The fuzzy background represents the Gaussian 
    white noise spread out through the spacetime.}
   \label{fig:RindlerAcoustic_Final2}
\end{figure}

Once we have fixed the criteria to define a classical analog of the Minkowski vacuum, we must ask 
a uniformly accelerated observer in the analog spacetime to extract the two and four-point functions 
and compare them with the corresponding quantum ones. The worldline of a uniformly accelerated 
observer with constant proper analog acceleration $A$ is
\begin{eqnarray}
   vt &=& (v^2/A)\sinh({A\tau/v}), \nonumber\\
   z &=&  (v^2/A)\cosh({A\tau/v}), \nonumber
   \label{equ:rindler}
\end{eqnarray} 
where $ \tau $ is the analog proper time (i.e., the length of the trajectory in the analog spacetime). 
The parameter $t$ is the time measured in the laboratory frame and can be seen to rapidly increase with $\tau$. 
In  Fig.~(\ref{fig:RindlerAcoustic_Final2}), the wordline of such an observer is exhibited in both the Minkowski and 
analog spacetimes. 

Along the observer's trajectory, $z^2 - v^2t^2 = v^4/A^2$, the field $ \mathcal{A}(t,z) $ can be Fourier analyzed
with respect to the Rindler frequency $\omega$ as
\begin{equation} \label{normalmode}
    \tilde{\mathcal{A}}(\omega) = \int_{-\tau_M}^{\tau_M} d\tau \mathcal{A}(t,z) e^{i \omega \tau},
\end{equation} 
where $\pm \tau_M = \pm (v/A) \, \operatorname{arcosh} [LA/(2 v^2)]$ are the analog proper instants 
when the observer's ride starts and finishes.   

Firstly, let us compute the two-point correlations between different modes as measured by Rindler observers.
By using Eq.~(\ref{normalmode}) with Eq.~(\ref{Afield}) and imposing 
Eqs.~(\ref{equ:clascorr1})-(\ref{equ:clascorr3}), we find 
$$
\left\langle \tilde{\mathcal{A}}(\omega_i) \tilde{\mathcal{A}}^{*}(\omega_j) \right\rangle 
\equiv
\mathcal{C}_2(\omega_i, \omega_j)  
$$
with
\begin{equation}
    \mathcal{C}_2(\omega_i, \omega_j) 
    \! = \!\! \sum_{m=1}^{N}  \frac{I}{2 m \pi} [f_m(\omega_i)f_m(\omega_j)  +  f_m(-\omega_i)f_m(-\omega_j)] 
    \label{equ:cor2}
\end{equation} 
where 
\begin{align}
f_m(\omega) = 
\begin{cases} 
    (-1)^{m/2} C_m(\omega), & \text{for even}\; m \\
    (-1)^{(m+1)/2} S_m (\omega), & \text{for odd}\; m 
\end{cases}
\label{equ:fm}
\end{align} 
and
\begin{eqnarray*}
   C_m(\omega) &=& \int_{-\tau_M}^{\tau_M} d\tau \, \cos\left(\frac{m \pi v^2}{A L}{e}^{-A \tau/v} + \omega \tau\right) \\
   S_m(\omega) &=& \int_{-\tau_M}^{\tau_M} d\tau \, \sin\left(\frac{m \pi v^2}{A L}{e}^{-A \tau/v} + \omega \tau\right).
\end{eqnarray*} 
Equation~(\ref{equ:cor2}) is what experimentalists should measure. (Leonhard et 
al.~\cite{Leonhardt2018} carried out their experiment taking into account a single mode rather than white noise. 
Moreover, they considered the field $\mathcal{A} (t,z)$ to be fixed at the left 
wall, while we have adopted boundary conditions~(\ref{modesboundaryconditions}) in compliance 
with the laws of hydrodynamics~\cite{lamb2015hydrodynamics}.) 
\begin{figure}
   \centering
   \includegraphics[width=85mm]{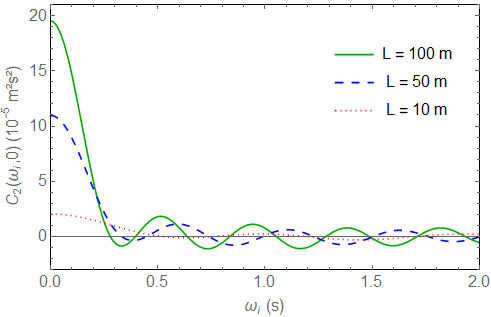}
    \caption{$ \mathcal{C}_2(\omega_i, 0) $ as a function of $ \omega_i$ for $ L = 10 $~{\rm m}, $ 50$~{\rm m} and $ 100 $~{\rm m}. 
    The parameters used in this plot are $ A = 0.1~{\rm m/s^2}$, $h = 0.01~{\rm m}$, $ v = 0.3~{\rm m/s} $, 
    and $ I = 10^{-6}~{\rm m^2}$.  $N = L/(100 \pi h)$ to guarantee  condition~(\ref{equ:condition}). }
   \label{fig:prediction}
\end{figure}

In Fig.~(\ref{fig:prediction}), we show  $\mathcal{C}_2(\omega_i,0)$ obtained by an observer with 
$A = 0.1~ {\rm m/s^2}$ assuming  $ h = 0.01~{\rm m}$  and $ L = 10~{\rm m}$, $50~{\rm m}$, 
and $100~{\rm m}$. We note that $ L \gg v^2/A $, where $ v = \sqrt{g h} \approx 0.3~{\rm m/s}$, i.e.,  
the length of the water basin is 
large compared to the other length scales of the problem in order to ensure that the walls have a relatively 
small impact on the system and  allow for the traces of the Unruh effect to become more apparent. 
For $ L = 10~{\rm m}$, $ 50~{\rm m}$, and $ 100~{\rm m}$, the duration of the experiment is $15.7~{\rm s}$, 
$80~{\rm s}$, and $160~{\rm s}$, respectively. 

Now, it is useful to compare the results above with the one obtained when $ L \to \infty $.
For this purpose, let us first cast Eq.~(\ref{equ:cor2}) as
\begin{equation}
\begin{aligned}
    \mathcal{C}_2(\omega_i, \omega_j) = \sum_{\substack{m=2 \\ m\, {\rm even}}}^{N} & \frac{I}{m \pi} [C_m(\omega_i)C_m(\omega_j) \\ 
    + & C_m(-\omega_i)C_m(-\omega_j)] \\ 
    + \sum_{\substack{m=1 \\ m\, {\rm odd}}}^{N} & \frac{I}{m \pi} [S_m(\omega_i)S_m(\omega_j) \\
    + & S_m(-\omega_i)S_m(-\omega_j)].
    \label{equ:cor21}
\end{aligned}
\end{equation}
In the $L\to \infty$ limit, we have 
\begin{equation*} 
 \frac{2 \pi}{L}\sum_{\substack{m=2 \\ m\, {\rm even}}}^{N}, \quad
 \frac{2 \pi}{L}\sum_{\substack{m=1 \\ m\, {\rm odd}}}^{N} \to \int_{0}^{\infty} dk 
 \end{equation*}
and
\begin{eqnarray*}
   C_m (\omega) \to C_k(\omega) 
   &=& \int_{-\infty}^{\infty} d\tau \, \cos\left(\frac{k v^2}{A}{e}^{-A \tau/v} + \omega \tau\right), \\
   S_m (\omega) \to S_k(\omega) 
   &=& \int_{-\infty}^{\infty} d\tau \, \sin\left(\frac{k v^2}{A}{e}^{-A \tau/v} + \omega \tau\right).
\end{eqnarray*}
Then, by using~\cite{Mathematica} 
\begin{eqnarray}
 \int_{-\infty}^{\infty} d\eta \exp(\pm i \beta e^{\mp \eta} + i \varpi \eta) 
 &=& - \beta^{\pm i \varpi} e^{\pi \varpi/2} \Gamma(\mp i \varpi) 
 \nonumber \\
 \int_{-\infty}^{\infty} d\eta \exp(\mp i \beta e^{\mp \eta} + i \varpi \eta) 
 &=& - \beta^{\pm i \varpi} e^{-\pi \varpi/2} \Gamma(\mp i \varpi) 
 \nonumber
 \label{equ:gamma1} 
\end{eqnarray}
and
\begin{equation*}
 |\Gamma(i \eta)|^2 = \pi/[\eta \sinh(\pi \eta)], \quad \eta \in \mathbb{R}
\end{equation*}
we obtain    
\begin{equation}\label{equ:corUnruh}
    \mathcal{C}_2(\omega_i,\omega_j) 
    =\frac{2 \pi I}{\omega_i} \left(\frac{1}{2} + \frac{1}{e^{2 \pi \omega_i v/A}-1}\right) \delta(\omega_i - \omega_j). 
    \end{equation} 
We see that the curves in Fig.~\ref{fig:prediction} are consistent with Eq.~(\ref{equ:corUnruh}) in the sense that 
the larger the $L$, the sharper the peaks around $\omega_i=0$ are.

\section{Connection with the Unruh effect} \label{sec:Unruheffect}

In order to see how Eq.~(\ref{equ:corUnruh}) connects with the Unruh effect, let us 
perform the corresponding calculation in the spacetime $ (\mathbb{R}^2,\eta)$
considering a quantum free massless scalar field
\begin{equation} \label{equ:quantfield2}
\hat{\Phi}(x^{\mu}) = \int_{-\infty}^{+\infty} dk 
\left (\frac{\hbar c^2}{4 \pi \varpi_k}\right)^{1/2}
[\hat{a}_k e^{i k_\mu x^{\mu}} + \hat{a}^{\dagger}_k e^{-i k_\mu x^{\mu}}],
\end{equation}
where $[\hat a_k, \hat a_{k'}^\dagger] = \delta (k-k')$. 
Equation~(\ref{equ:quantfield2}) can be straightforwardly otained from   
Eq.~(\ref{equ:quantfield}) under the identifications:
$$
\frac{2 \pi}{L} \sum_{\substack{m=-\infty \\ m \neq 0}}^{\infty} \to \int_{-\infty}^{+\infty} dk, \; 
\frac{2 \pi m}{L} \to k, \;
\sqrt{\frac{L}{2\pi} } a_m \to a_k.
$$

Now, let us take 
\begin{equation*}
     \widetilde{\hat{\Phi}}(\omega) = \int_{-\infty}^{\infty} d\tau \, \hat{\Phi}(t,z) e^{i \omega \tau}.
\end{equation*}
to be the Fourier transform of $ \hat{\Phi} (t,z) $ along the unextendible worldline of a uniformly accelerated observer 
with acceleration $ a $:
\begin{eqnarray} 
   ct &=& (c^2/a)\sinh({a\tau/c}), \nonumber \\
   z &= & (c^2/a)\cosh({a\tau/c}). \nonumber
\end{eqnarray}
  
 Then, following last section calculations, we obtain
\begin{eqnarray}
   \mathcal{Q}_2(\omega_i, \omega_j) 
   &\equiv& 
   \left\langle 0 \left| \widetilde{\hat{\Phi}} (\omega_i)  \widetilde{\hat{\Phi}}^{\dagger} (\omega_j)  
+ \widetilde{\hat{\Phi}}^{\dagger} (\omega_j) \widetilde{\hat{\Phi}} (\omega_i) \right|0 \right\rangle /2, 
\nonumber \\
  & = &\frac{2 \pi \hbar c^2}{\omega_i} \left(\frac{1}{2} + \frac{1}{e^{2 \pi \omega_i c/a}-1}\right) \delta(\omega_i - \omega_j)
 \nonumber \\
    \label{equ:quantcorr}
\end{eqnarray} 
The similarity between Eqs.~(\ref{equ:corUnruh}) and~(\ref{equ:quantcorr}) is clear. 
In particular, $ I $ in Eq.~(\ref{equ:corUnruh}) plays the role of $ \hbar c^2$ in 
Eq.~(\ref{equ:quantcorr})~\cite{units}. This is particularly interesting, since the value of 
the strength $ I $ can be easilly controlled by the experimentalist. Furthermore, just as in the 
quantum case, a Planckian term appears in Eq.~(\ref{equ:corUnruh}). From a quantum 
perspective, the thermal distribution is characterized by the  $ [\exp(E/k_B T)-1]^{-1} $ term, 
where $ E = \hbar \omega $ is the energy of a particle with angular frequency $\omega $. 
In the classical case, however, the energy of a surface wave is not proportional to $ \omega $, 
making it impossible to obtain a corresponding physical temperature from Eq.~(\ref{equ:corUnruh}). 
We can, nonetheless, formally define an analog temperature, 
\begin{equation}
   T_A = \frac{\hbar A}{2 \pi k_B v},
    \label{equ:tempclas}
\end{equation}
as a parameter which characterizes the Planckian distribution of the correlation function.
  
\section{Discriminating among distinct noises} \label{sec:othernoises} 

\begin{figure}
\includegraphics[width=1\linewidth]{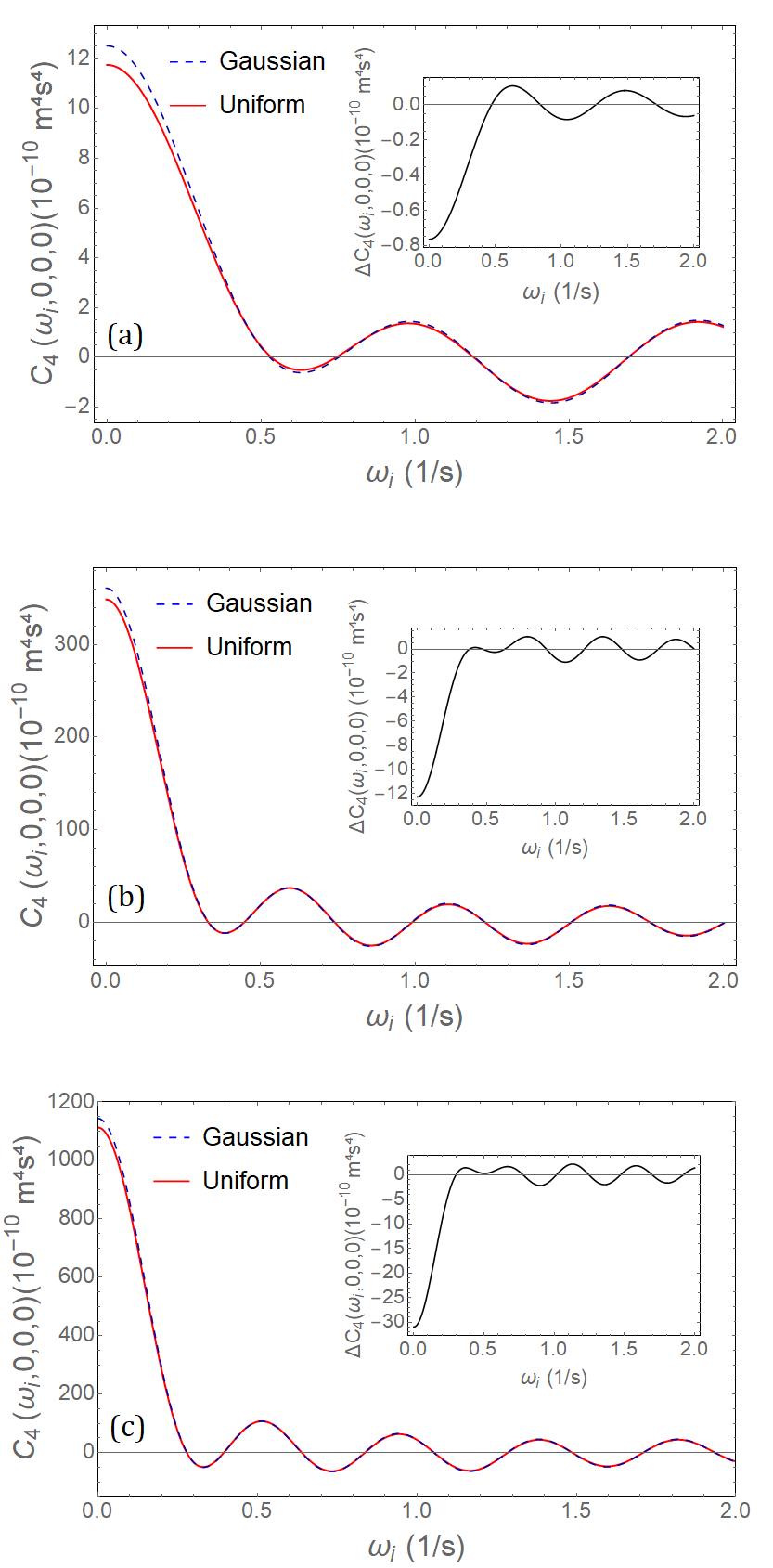}
\caption{The four-point functions for the uniform (solid line) and Gaussian (dashed line) distributions are
plotted. The upright inserted graphs show the difference between them. The first, second and 
third plots assume $L=10~{\rm m}$, $L=50~{\rm m}$, and $L=100~{\rm m}$, respectively. }
  \label{fig:4point}
  \end{figure}

As discussed at the end of Sec.~\ref{Sec:AnalogVac}, although distinct white noises, $P(r_m)$, will lead to the 
same two-point functions  $\mathcal{C}_2(\omega_i,\omega_j)$,  they will differ, in general, for higher-order ones. 
In this section, we compare  $\mathcal{C}_4(\omega_i,\omega_j,\omega_k,\omega_l) $ obtained assuming 
Gaussian~(\ref{gauss}) and uniform~(\ref{uniform}) distributions. This should give a feeling on how much 
our results above may be sensitive to deviations from Gaussianity as one prepares the classical ``vacuum'' 
state in the laboratory. We begin by computing the four-point momenta 
$ \left\langle r_i r_{j} r_{k} r_{l}\right\rangle $, where $ r_i = p_i, \,q_i $.
The only nonvanishing ones are  
\begin{eqnarray*}
     \left\langle p_i p_{j} p_{k} p_{l}\right\rangle 
     &=& \frac{3 I^2}{4}\delta_{(i j}\delta_{k l)} - \frac{3}{10}\alpha I^2 \delta_{i j}\delta_{j k}\delta_{k l}, 
     \nonumber \\
    \left\langle q_i q_{j} q_{k} q_{l}\right\rangle 
     &=& \frac{3 I^2}{4}\delta_{(i j}\delta_{k l)} - \frac{3}{10}\alpha I^2 \delta_{i j}\delta_{j k}\delta_{k l}, 
     \nonumber \\
     \left\langle q_i q_{j} p_{k} p_{l}\right\rangle &=& \frac{I^2}{4} \delta_{ij}\delta_{kl},
\end{eqnarray*}
where $\alpha=0$ and  $\alpha=1$ for Gaussian and uniform distributions, respectively. 
Using it, we obtain 
\begin{eqnarray}
 \mathcal{C}_4(\omega_i,\omega_j,\omega_k,\omega_l) 
 &\equiv &
 \left\langle 
 \widetilde{\mathcal{A}}(\omega_i)
 \widetilde{\mathcal{A}}(\omega_j) 
 \widetilde{\mathcal{A}}^{*}(\omega_k) 
 \widetilde{\mathcal{A}}^{*}(\omega_l) 
 \right\rangle
 \nonumber \\
 &= & \left( \frac{I}{2\pi}\right)^2\sum_{m,n=1}^{N} \frac{1}{m n}
 \nonumber \\
 & \times & [g_{mnmn}(\omega_i, \omega_j,\omega_k, \omega_l) 
 \nonumber \\
 &+ & g_{mnnm}(\omega_i, \omega_j,\omega_k, \omega_l)
 \nonumber \\
 &+ & g_{mmnn}(\omega_i, -\omega_j,\omega_k, -\omega_l)
 \nonumber\\  
 & +& g_{mnmn}(\omega_i, -\omega_j,\omega_k, -\omega_l)
 \nonumber \\
 &+ & g_{mmnn}(\omega_i, -\omega_j,-\omega_k, \omega_l) 
  \nonumber\\ 
 &+ & g_{mnnm}(\omega_i, -\omega_j,-\omega_k, \omega_l)] 
 \nonumber \\
 & - & \frac{3\alpha}{10} \left(\frac{I}{2\pi}\right)^2 \sum_{m=1}^{N} \frac{1}{m^2}
 \nonumber\\ 
 &\times & [g_{mmmm}(\omega_i,\omega_j,\omega_k,\omega_l) 
 \nonumber \\
 & + & g_{mmmm}(\omega_i,-\omega_j,\omega_k,-\omega_l)
  \nonumber\\ 
 &+ & g_{mmmm}(\omega_i,-\omega_j,-\omega_k,\omega_l)],
\end{eqnarray} 
where
\begin{eqnarray}
 g_{nmn'm'}(\omega_i,\omega_j,\omega_k,\omega_l) 
 & = & [f_n(\omega_i)f_m(\omega_j)f_{n'}(\omega_k)f_{m'}(\omega_l)
  \nonumber\\ 
 &+ & f_{n}(-\omega_i)f_m(-\omega_j)f_{n'}(-\omega_k) \nonumber \\
 &\times & f_{m'}(-\omega_l)]
\end{eqnarray} 
and $ f_m (\omega_i) $ is given in Eq.~(\ref{equ:fm}). In Fig.~\ref{fig:4point}, 
we plot $ \mathcal{C}_4(\omega_i,0,0,0)$ 
for the Gaussian and uniform cases assuming $L = 10~{\rm m}, 50~{\rm m}$, and $100~{\rm m}$. 
The difference between them, albeit small, is still noticeable, as shown in the inserted plots. 

For the sake of completeness, let us finally exhibit $ \mathcal{C}_4(\omega_i,\omega_j,\omega_k,\omega_l)$
and its corresponding quantum counterpart $\mathcal{Q}_4(\omega_i,\omega_j,\omega_k,\omega_l)$
in the limit $ L \to \infty $ for the Gaussian case. By following the very same procedures as in 
Sec.~\ref{sec:2point} and~\ref{sec:Unruheffect}, we obtain the following results:
\begin{eqnarray}
    && \mathcal{C}_4(\omega_i,\omega_j,\omega_k,\omega_l)  
    = \frac{4 I^2 \pi^2}{\omega_i \omega_j\tanh(\pi \omega_i v/a)  \tanh(\pi \omega_j v/a)}
    \nonumber \\ 
    && \times [\delta(\omega_i - \omega_k)\delta(\omega_j - \omega_l) 
    + \delta(\omega_i - \omega_l)\delta(\omega_j - \omega_k)]
    \nonumber 
\end{eqnarray} 
and
\begin{eqnarray}
    && \mathcal{Q}_4(\omega_i,\omega_j,\omega_k,\omega_l) 
    = \frac{4 \hbar^2 c^4 \pi^2}{\omega_i \omega_j \tanh(\pi \omega_i v/a)  \tanh(\pi \omega_j v/a)}
    \nonumber \\ 
    && \times [\delta(\omega_i - \omega_k)\delta(\omega_j - \omega_l) 
    + \delta(\omega_i - \omega_l)\delta(\omega_j - \omega_k)],
\end{eqnarray} 
 where 
$$
\mathcal{Q}_4(\omega_i,\omega_j,\omega_k,\omega_l) \equiv 
\left\langle 0 \left| 
{\cal S }
\left[ 
\widetilde{\hat{\phi}} (\omega_i) 
\widetilde{\hat{\phi}} (\omega_j)
\widetilde{\hat{\phi}}^{\dagger} (\omega_k) 
\widetilde{\hat{\phi}}^{\dagger}(\omega_l) 
\right]
 \right| 0  \right\rangle 
$$
and ${\cal S}$ is the total symmetrization operator as required by the procedure explained in Sec.~\ref{qvacuum}. 

\section{Conclusions} \label{sec:conclusions} 

We have established a way to mimic some aspects of the quantum vacuum of a massless free scalar field 
with classical gravity waves. Then, we have calculated the two- and four-point functions of the Fourier 
transform of the classical field along a uniformly accelerated wordline of the analog spacetime, taking into account
the boundary conditions as dictated by hydrodynamics. We have shown how to link the two- and four-point 
functions with the Unruh effect in the limit where the water basin is large enough. Furthermore, we have investigated
how deviations from Gaussianity in the choice of the classical ``vacuum'' may impact in the process of producing a 
``faithful'' classical analog of the Unruh effect.

\acknowledgments
G.~B. and J.~R. ackowledge full support from Coordena\c c\~ao de Aperfei\c coamento de Pessoal 
de N\'ivel Superior (Capes) under grant No.~88882.330762/2019-01 and S\~ao Paulo 
Research Foundation (FAPESP) under grant No. 2017/26809-1, respectively. 
A.~L. and G.~M. are grateful to FAPESP under Grant No. 2017/15084-6 and Conselho 
Nacional de Desenvolvimento Cient\'\i fico e Tecnol\'ogico  under grant No. 301544/2018-2, 
respectively, for partial support. 


\begin{thebibliography}{99}
\bibitem{units} 
The units differ because our field~(\ref{Afield0}) has unit of length, while field~(\ref{equ:quantfield2}) has the same 
unit of the electromagnetic potential: ${\rm g^{1/2} \,cm^{1/2}/ s} $ in CGS.
\end{thebibliography}
\end{document}